\documentclass[lettersize,journal]{IEEEtran}
\usepackage{amsmath,amsfonts}
\usepackage{algorithmic}
\usepackage{algorithm}
\usepackage{array}
\usepackage{float}
\usepackage[caption=false,font=normalsize,labelfont=sf,textfont=sf]{subfig}
\usepackage{multirow}
\usepackage[table,xcdraw]{xcolor}
\usepackage{textcomp}
\usepackage{stfloats}
\usepackage{url}
\usepackage{verbatim}
\usepackage{graphicx}
\usepackage{cite}
\usepackage{booktabs}
\usepackage{setspace}  % 导入 setspace 宏包
\usepackage[colorlinks, citecolor=blue]{hyperref}
\begin{document}
\title{Dynamic Perturbation-Adaptive Adversarial Training on Medical Image Classification}
\author{Shuai Li, Xiaoguang Ma, \IEEEmembership{Member, IEEE}, Shancheng Jiang, and Lu Meng
\thanks{S. Li, X. Ma, and L. Meng are with the College of Information Science and Engineering, Northeastern University, Shenyang 110819, China. (e-mail: 2270882@stu.neu.edu.cn, maxg@mail.neu.edu.cn, menglu1982@gmail.com) S. Li and X. Ma are with the Foshan Graduate School, Northeastern University, Foshan 528311, China. X. Ma is with the Faculty of Robot Science and Engineering, Northeastern University, Shenyang 110819, China, and the State Key Laboratory of Synthetical Automation for Process Industries, Northeastern University, Shenyang 110819, China. S. Jiang is with the School of Intelligent Systems Engineering, Sun Yat-Sen University, No. 135, Xingang Xi Road, Guangzhou, PR China, and Guangdong Provincial Key Laboratory of Fire Science and Technology, Guangzhou 510006, China.(e-mail: jiangshch3@mail.sysu.edu.cn)}}

\maketitle

\markboth{}{Dynamic Perturbation-Adaptive Adversarial Training on Medical Image Classification}


\maketitle

\begin{abstract}
Remarkable successes were made in Medical Image Classification (MIC) recently, mainly due to wide applications of convolutional neural networks (CNNs). However, adversarial examples (AEs) exhibited imperceptible similarity with raw data, raising serious concerns on network robustness. Although adversarial training (AT), in responding to malevolent AEs, was recognized as an effective approach to improve robustness, it was challenging to overcome generalization decline of networks caused by the AT. In this paper, in order to reserve high generalization while improving robustness, we proposed a dynamic perturbation-adaptive adversarial training (DPAAT) method, which placed AT in a dynamic learning environment to generate adaptive data-level perturbations and provided a dynamically updated criterion by loss information collections to handle the disadvantage of fixed perturbation sizes in conventional AT methods and the dependence on external transference. Comprehensive testing on dermatology HAM10000 dataset showed that the DPAAT not only achieved better robustness improvement and generalization preservation but also significantly enhanced mean average precision and interpretability on various CNNs, indicating its great potential as a generic adversarial training method on the MIC.
\end{abstract}

% \doublespacing 
\section{Introduction}
\label{introduction}
\IEEEPARstart{M}{edical} image approximately accounted for 90\% of all data in healthcare \cite{zhou2021review}, providing vital evidence for clinical diagnosis and interventions. Attribute to rapid improvement of deep learning (DL) algorithms, which could automatically extract stereoscopic and abstract lesion features, end-to-end learning and identification of Medical Image Classification (MIC) developed quickly, wherein heavy workload of doctors could be reduced. Among the DL algorithms, convolutional neural networks (CNNs) showed impressive abilities of feature extraction and provided unique and powerful capabilities in various MIC tasks, especially for skin \cite{datta2021soft}, lung \cite{anthimopoulos2016lung}, and brain cancer \cite{pereira2016brain}. In fact, skin cancer investigation accounted for 33\% of all analyzable tumors work \cite{zunair2020melanoma}, and existing technologies using CNNs could achieve accuracy over 87\% for skin lesion classification \cite{bratchenko2022classification},\cite{yao2021single}.

However, \cite{szegedy2013intriguing} elucidated that injecting small perturbations into raw image data, could easily fool the DL networks, raising great concerns on models' robustness and seriously hindering its applications \cite{madry2017towards}. Therefore, more research had been focusing on obtaining satisfactory robustness, including filtering out perturbations on inputs \cite{zorzi2016robust} and detecting adversarial perturbations before inputs entering the DL networks \cite{grosse2017statistical}. Specifically, augmenting training data with adversarial examples (AEs), i.e., adversarial training (AT) \cite{madry2017towards}, was found to be an effective robustness improvement method. In fact, research on the MIC showed that various methods based on the AT methods had achieved state-of-the-art robustness performance \cite{mahmood2018unsupervised},\cite{yoo2022noise}. However, these AT-based methods rarely considered feature exploration of raw data during adversarial training, greatly wasting MIC resources. In \cite{goodfellow2014explaining}, a standard adversarial training (SAT) method, wherein the raw data and AEs jointly participated in the training, was proposed to improve the robustness and became a mainstream AT approach.  

Nevertheless, both the AT and SAT suffered from two significant drawbacks, i.e., heavy computational overhead and generalization decline \cite{tsipras2018robustness}. Although the computational overhead of the AT methods could be 2 to 10 times higher than that of standard training, it could be reduced by simplifying redundant structures with lightweight networks. However, the generalization decline was more challenging to overcome. \cite{ma2022adaptive} elucidated that fixed perturbation sizes in the SAT could cause the generalization decline, and proposed an adaptive-margin adversarial training (AMAT) for the MIC, wherein a loss-defined-margin strategy was used to craft adaptive AEs with various perturbation sizes. However, in the AMAT, thresholds that constituted the recognition criteria were transferred from a well-trained external network. Moreover, they were fixed and static, and could not be updated during training.

In fact, the external transference ignored possible static constricts during perturbation searching, easily leading to cases where only either the robustness or the generalization could be improved. Moreover, training effectiveness of traditional AT methods greatly relied on the quality and effectiveness of apriori knowledge from external transference. In addition, in the field of computer vision, improvement of either the generalization or the robustness, or even both, did not guarantee to bring better interpretability, which was greatly preferable for the MIC, especially when using various CNNs \cite{salih2023explainable}.

In this paper, we illustrated a data-level understanding among robustness and generalization on the MIC, and proposed a dynamic perturbation-adaptive adversarial training (DPAAT) method, making threshold transference dynamically updated to help generate more suitable perturbation sizes, by collecting loss information as reference knowledge of the training. Our main contributions were summarized as follows:
\begin{itemize}
\item A loss distribution property $\Delta L_{avg}$ was collected as reference knowledge of data separation, i.e., relatively fragile and relatively stable data, through dynamically determining loss central tendency. For relatively fragile data, their perturbation sizes were decreased to avoid classifier boundary crossing-over, and more perturbations were supplemented to facilitate the robustness improvement of relatively stable data.  
\item We proposed a dynamic perturbation-adaptive adversarial training (DPAAT) method, which placed adversarial training in a dynamic learning environment to generate more suitable adaptive perturbation sizes. Meanwhile, we optimized synchronization between the robustness and generalization with in-depth feature explorations of the raw data.
\item We experimentally confirmed significant improvements of the DPAAT on the dermatology HAM10000 dataset. In robustness testing, six CNNs using the DPAAT outperformed those using AT, SAT, and AMAT under iterative PGD and IFGSM attacks. Meanwhile, the DPAAT obtained the best generalization preservation accuracy, i.e., the closest to that of standard training.
\item The DPAAT also got superior interpretability of the CNNs over standard and the AT methods, i.e., covering more coherent and accurate lesion regions in most of the Grad-CAM images.
\end{itemize}
\section{Related works}
\label{relatedwork}
\subsection{Review of Adversarial Training Methods}
\label{Section 2.1}
We defined $S=\left\{ {(x_i,y_i)} \right\}^{n}_{i=1}$ as training datasets, wherein $n$ represented numbers of training data and $(x_i,y_i)$ was group of the $i_{th}$ raw data $x$ and corresponding true label $y$. Adversarial examples $x^{adv}$ were examples within perturbation balls around the raw data $x$. The perturbation balls were defined as follows,
\begin{align}
\label{Eq1}
B \sim B(p,x,\epsilon) = \{ x_i^{adv}|{\left\| {x_i^{adv} - {x_i}} \right\|_p} \le \epsilon \} _{i = 1}^n,  
\end{align}
where ${\left\| {\cdot} \right\|_p}$ represented $l_{p}$-norm, which was used to calculate pixel-pixel distance between $x$ and $x^{adv}$ and $p$ was an optional parameter to determine calculation method of ${\left\| {\cdot} \right\|_p}$. Pre-set perturbation size $\epsilon$ represented perception disparity between $x$ and $x^{adv}$, deciding size of the perturbation balls $B$, and even a tiny $\epsilon$ was enough to cause a large number of false predictions of the DL models for $x^{adv}$. During AT processes, $x$ would be deliberately mixed with some small perturbations, and the DL models would become more robust to $x^{adv}$ by adapting the perturbations. \cite{madry2017towards} made a unified statement for the AT, i.e.,
\begin{align}
\label{Eq2}
    \mathop {\min }\limits_{{\theta}} {E_{(x,y) \in {S}}}[{\mathop {\max }\limits_{x^{adv}  \in B} L(\theta ,x^{adv} ,y)}],
\end{align}
where $L(\cdot)$ represented loss function of standard training and Cross-Entropy function was selected in this paper, $\theta$ represented parameters of the DL models, and $E(\cdot)$ was empirical risk minimization strategy. The AT statement consisted of two parts, i.e., internal maximization to find $x^{adv}$ causing the largest losses and external minimization to find $\theta$ to minimize the high losses. The traditional AT \cite{madry2017towards} used a loss function from standard training, i.e.,
\begin{align}
\label{Eq3}
L_{AT} = L(\theta ,{x^{adv}},y).
\end{align}
$x^{adv}$ was usually made intentionally and $x$ was not considered in the training. \cite{goodfellow2014explaining} showed that the DL models could be regularized by considering both $x^{adv}$ and $x$ during training, wherein the SAT could be illustrated as follows,
\begin{align}
\label{Eq4}
\mathop {\min }\limits_{{\theta}}& {E_{(x,y) \in {S}}}[{\mathop {\max }\limits_{x^{adv} \in B} {L_{SAT}}}],\nonumber
    \\{L_{SAT}} = & \alpha L({\theta},{x^{adv}},y) + (1 - \alpha )L({\theta},x,y),    
\end{align}
where $\alpha$ was a hyper-parameter to balance speed and magnitude of the loss convergence between $x^{adv}$ and $x$.\\
\indent In both the AT and SAT, perturbation sizes $\epsilon$ for all training data were fixed, and this was not effective for robustness improvement. Therefore, AMAT \cite{ma2022adaptive} was proposed to give various $\epsilon$ for each training data by a concept of loss margin, i.e.,
\begin{align} 
\label{Eq5}
{\epsilon_{AMAT}} =
\begin{cases}
{\epsilon + \Delta \epsilon},&{\text{if}}\ L_{AT} < \xi ,\\
{{\frac{1}{2}( \epsilon+ {{\left\| {{x^{adv}} - x} \right\|}_p})},}&{\text{otherwise,}}
\end{cases}
\end{align}
where $\epsilon _{AMAT}$ represented adaptive $\epsilon$, $\Delta \epsilon$ was a hyper-parameter to enlarge the loss margin, and $\xi$ was the threshold controlling perturbations in the AMAT. Moreover, new perturbation balls would be generated due to the $\epsilon_{AMAT}$, i.e.,
\begin{align}
\label{Eq6}
B_{AMAT}& \sim B(p,x,L_{ADV})
= 
\nonumber
\\&\{ x_i^{adv}| 
{{{\left \| {x_i^{adv} - {x_i}} \right\|}_p} \le {\epsilon _i},{\epsilon _i} = \epsilon _{AMAT}}\} _{i = 1}^n,
\end{align}
and the AMAT \cite{ma2022adaptive} was reflected as follows,
\begin{align}
\label{Eq7}
    \mathop {\min }\limits_{{\theta}} & {E_{(x,y) \in {S}}}[ {\mathop {\max }\limits_{x^{adv} \in {B_{AMAT}}} L_{AMAT}} ],\nonumber
\\{L_{AMAT}} &= \frac{1}{2}({L({\theta},{x^{adv}},y) + L({\theta},x,y)}).
\end{align}
\subsection{Review of Adversarial Attacks}
\label{Section 2.2}
\cite{szegedy2013intriguing} proposed a box-constrained attack method, i.e., Limited Memory Broyden-Fletcher-Goldfarb-Shanno (L-BFGS), where the CNNs were found to be vulnerable to $x^{adv}$. However, the L-BFGS was not suitable to be used as the training adversary for typical CNN backbones, due to its heavy nonlinear overhead. \cite{goodfellow2014explaining} proposed a fast gradient sign method (FGSM) as the training adversary to generate $x^{adv}$, i.e.,
\begin{align}
\label{Eq8}
   {x^{adv}} = x + \epsilon sign\left[ {{\nabla _x}L(\theta ,x,y)} \right],
\end{align}
where $sign(\cdot)$ was a symbolic function to return gradient signs. Although the DL models using the FGSM had better robustness on single-step attacks due to its local optimality in internal maximization, they could not defense iterative attacks. Instead, projected gradient descent (PGD) \cite{madry2017towards},  an iterative attack method, was proposed as the training adversary against more powerful malicious attacks, i.e.,
\begin{align}
\label{Eq9}
    x_{k + 1}^{adv} = \mathop {\Pr {\rm{oj}}}\limits_{x_{k}^{adv} \in B} \{ x_k^{adv} + \eta sign\left[ {{\nabla _{x_k^{adv}}}L(\theta ,x_k^{adv},y)} \right]\},
\end{align}
where $x^{adv}_0=x$, $x_k^{adv}$ was $x^{adv}$ of the $ k_{th}$ attack, $\eta$ was step size of the iterative attack, and ${\Pr {\rm{oj}}}\left\{  \cdot  \right\}$ was an operator that projected perturbations exceeding $\epsilon$, i.e., $\left\| {x_i^{adv} - {x_i}} \right\|_p >  \epsilon \ $, into the perturbation ball $B$. The PGD was a heuristic and strong first-order attack method, and could resist other attacks relying on first-order information provided that the DL models were robust to malicious attacks from the PGD.  
\subsection{Overview of Our Motivation}
\label{Section 2.3}
After the DL models obtained adversarial settings, their generalization accuracy, i.e., classification ability for $x$, usually dropped by 10\%, and the robustness, i.e., resistance to $x^{adv}$, could be improved by more than 45\% \cite{tsipras2018robustness}. Our motivation of this work was to improve robustness while preserving generalization as high as possible by generating adaptive perturbation sizes for each $x$. However, the perturbation size of each $x$ was always set to be the same and fixed \cite{ma2022adaptive} in conventional AT approaches. Before and after perturbation, the DL models usually predicted different results \cite{yang2022one}, i.e.,
\begin{align}
\label{Eq10}
    F: x &\to {x^{adv}},\nonumber\\
    \arg \max {G_f}(\theta ,F(x)) &\ne \arg \max {G_f}(\theta ,x),
\end{align}
where $F$ represented adversarial attacks mapping $x$ to $x^{adv}$. $G_{f}(\cdot)$ returned possibilities for all classes and $\arg \max(G_{f}(\cdot))$ was designed to give prediction results of the DL models. Training solely on $x$ made the DL models obtain classifier boundaries with optimal generalization (CBOG) for standard classification tasks using CNNs \cite{yang2022one}. However, alterations of the predicted results in \eqref{Eq10} revealed that deliberately crafted $x^{adv}$ could easily fool and cross over the CBOG, wherein generalization of the obtained classifier boundaries would become lower than that of the CBOG. Interestingly, some recent studies \cite{balaji2019instance},\cite{nie2021adaptive} pointed out that not all $x^{adv}$, as the same as those causing crossing-over phenomena, were deceptive for the CBOG. Instead, injecting more perturbations into these non-deceptive examples was beneficial to the robustness improvement. Unfortunately, this would not be allowed in conventional AT methods, since a fixed perturbation size prohibited perturbations beyond $\epsilon$. All the above discussion made it clear that fixed perturbation size $\epsilon$ hurt generalization ability and confined robustness improvement.

Practically, data far away from the CBOG were more robust, and ones closer to the CBOG were more vulnerable to malicious attacks \cite{yang2022one}, and the perturbation injection for each $x$ could be controlled by stabilizing the distance relationship between training data and the CBOG. Therefore, we divided all data into relatively fragile and relatively stable classes by considering the distance relationship between the training data and the CBOG. On the one hand, the relatively fragile data generated relatively large adversarial losses on the crafted $x^{adv}$, which were closer to the CBOG or even crossed over the CBOG, and the DL models trained with them could more likely cause generalization decline. On the other hand, the adversarial losses were smaller for relatively stable data, which were far away from the CBOG, and supplementing more perturbations on them could facilitate robustness improvements.

\section{Methodology}
\label{method}
\subsection{Synchronization loss}
\label{Section 3.1}
It was worth noting that our motivation was to improve robustness while minimizing generalization decline, thus synchronization between the two objectives was highly desired. However, in previous studies, $L_{SAT}$ was usually developed as a total loss function to update the DL models, while the synchronization was not fully considered, as shown in \eqref{Eq4} and \eqref{Eq7}. Moreover, model logits in this paper, i.e., outputs of the DL models after Softmax function was used, represented features information extracted from inputs. Thus, we used the model logits of $x$ and $x^{adv}$ to minimize synchronization losses. To avoid gradient disappearance during training, the synchronization loss could be represented by a symmetrical log-probability function, i.e.,
% \begin{align}
% \label{Eq11}
% L_{SYN}=
% \begin{cases}
% {L_{syn}},&{\text{if}}\ L_{syn} > 0 ,\\
% {0,}&{\text{otherwise,}}
% \end{cases}
% \end{align}
\begin{align}
L_{syn} = - \frac{1}{2}(y^{ori}&\log (y^{ori}/(y^{ori} + y^{adv})) \nonumber\\&
 + y^{adv}\log (y^{adv}/(y^{ori} + y^{adv})) )
\end{align}
where $y^{ori}$ and $y^{adv}$ were model logits from $x$ and $x^{adv}$, respectively. Obviously, smaller difference between $y^{ori}$ and $y^{adv}$ meant lower synchronization loss $L_{SYN}$. The entropy structures $L_{SYN}$ constrained the difference between $y^{ori}$ and $y^{adv}$ in feature extraction of the DL models. Thus, $y^{ori}$ and $y^{adv}$ achieved mutual supervision and perception alignment between $x$ and corresponding $x^{adv}$. After combining $L_{SYN}$, the total loss function of the DPAAT could be expressed by,
\begin{align}
\label{Eq12}
{L_{DPAAT}} = L_{SAT} +\beta L_{SYN},
\end{align}
where $\beta$ was a tuneable hyper-parameter.
\subsection{Dynamic Perturbation-Adaptive Adversarial Training}
\label{Section 3.2}
Referring to Support Vector Machine \cite {cortes1995support}, the distance relationship between the training data and classifier boundaries could be determined by loss calculation. In one epoch, all loss values of the training data could be collected into a dataset as follows,
\begin{align}
V=\left\{ {\Delta {L_i}{}|\Delta {L_i} = L(\theta ,x_i^{adv},y) - L(\theta ,x_i,y)} \right\}_i^n,
\end{align}
where $\Delta L_i$ represented adversarial loss changes, i.e., the difference between classification loss $ L(\theta,{x_i},y)$ and adversarial loss $L(\theta,x_i^{adv},y)$, caused by injecting perturbations into $x_i$. Therefore, loss distribution property could be used to mathematically illustrate data separation, i.e.,
\begin{align}
\label{Eq14}
\Delta {L_{avg}} = \frac{1}{m}\sum\limits_{i = 1}^m {\Delta {L_i}},
\end{align}
where $m$ represented size of a mini-batch and $\Delta L_{avg}$ was the distribution mean of the dataset $V$, reflecting central loss tendency of all $\Delta L$ in the mini-batch. 
\begin{figure*}[!t]
    \centering  \includegraphics[width=0.75\textwidth]{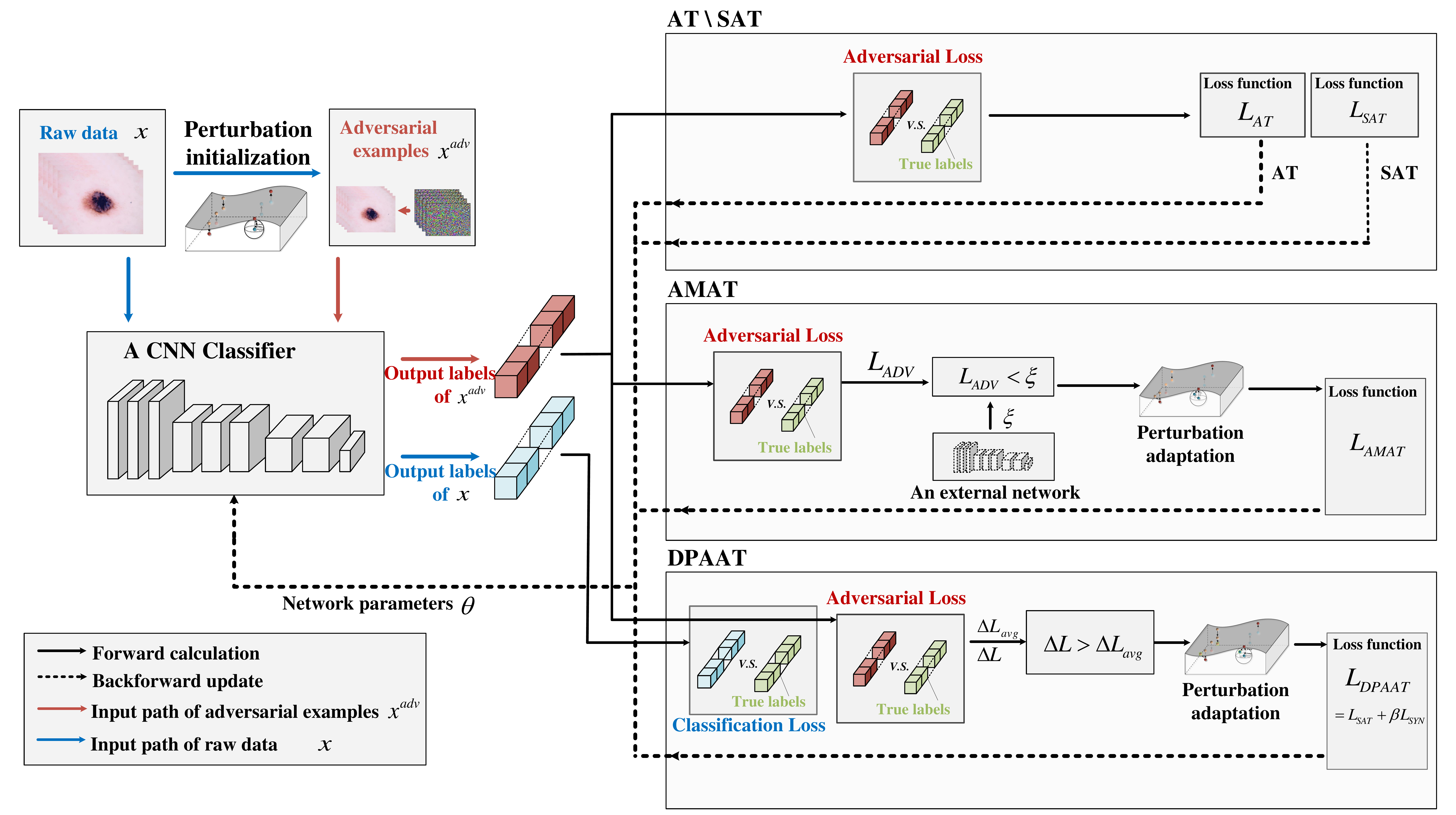}     \setlength{\belowcaptionskip}{0pt}
    \caption{The framework of the DPAAT and three conventional AT methods. Red and blue arrows were input paths of raw data $x$ and AEs $x^{adv}$, respectively. Solid black arrows represented forward calculation of input data and dotted black arrows represented back update of loss functions. Adversarial loss and classification loss were the loss values calculated by $L_{AT}$ on $x^{adv}$ and $x$, respectively.} 
    \label{fig:2-fra}
\end{figure*}

Specifically, the distance relationship had greater adversarial loss changes from $x$ to $x^{adv}$ when $\Delta L > \Delta L_{avg}$, indicating that corresponding data were relatively more fragile. On the contrary, $\Delta L$ of relatively stable data was smaller since the loss values changed only slightly and were more stable. Thus, the perturbation adaptation could be introduced as follows,
\begin{align}
\label{Eq15}
{\epsilon_{DPAAT}} &=
\begin{cases}
    {{{\left\| {{x^{adv}} - x} \right\|}_p}-\gamma}&{\text{if}}\ \Delta L > \Delta {L_{avg}},\\
    {{{\left\| {{x^{adv}} - x} \right\|}_p}+\gamma}&{\text{otherwise,}}
\end{cases}\nonumber \\
 \gamma  &= \epsilon \cdot (\left| {\Delta L - \Delta {L_{avg}}} \right|/\Delta {L_{avg}}),
\end{align}
where $\gamma$ was an adaptive length during adaptation processes. Obviously, the criteria of $\Delta L >\Delta L_{avg}$ helped pick out relatively fragile data in the mini-batches to decrease their perturbation sizes, i.e., cutting off $\gamma$. Meanwhile, perturbation sizes of the relatively stable data were increased, i.e., increasing $\gamma$. For $\Delta L$ close to the central loss tendency, the length $\gamma$ was tiny. Instead, the length $\gamma$ could be large, making all $\Delta L$ get close to $\Delta L_{avg}$ and preventing the perturbation size from being overly expanded or shrank. As loss information in $V$ was iteratively updated, varying $\Delta L_{avg}$ practically created a dynamic learning environment for perturbation injection. Therefore, the total training statement of the DPAAT could be reflected as follows,
\begin{align}
\label{Eq16}
    B_{DPAAT}& \sim B(p,x,\Delta L,\Delta L_{avg}) = \nonumber\\&
    \{ x_i^{adv}|\left\|{x_i^{adv} - {x_i}} \right\|_p\le \epsilon_i ,\epsilon_i = \epsilon_{DPAAT} \} _{i = 1}^n,\nonumber\\
    \mathop {\min }\limits&_ \theta{E_{(x,y) \in {S}}}[{\mathop {\max }\limits_{{x^{adv}} \in {B_{DPAAT}}} {L_{DPAAT}}}].
\end{align}
In summary, behaviors of adjusting the perturbation size of each data originated from self-transference of loss information knowledge in the DPAAT. This made each training data have its own perturbation size $\epsilon_{DPAAT}$ instead of a fixed perturbation size as shown in Alg. \ref{alg}. 
\begin{algorithm}[t]
	\caption{\footnotesize{Dynamic Perturbation-Adaptive Adversarial Training}}
	\label{alg}
	\begin{algorithmic}[1]
        \footnotesize
        \renewcommand{\algorithmicrequire}{\textbf{Input:}}
        \REQUIRE {$S$ represented the training dataset;
$i$ represented ID of $i_{th}$ data in $S$; 
$\epsilon$ represented fixed perturbation size; 
$\epsilon_{DPAAT}$ represented adaptive perturbation sizes; 
$\theta$ were initialized from pre-trained values on ImageNet dataset.}
        \renewcommand{\algorithmicensure}{\textbf{Output:}} 
        \ENSURE {a robust model with model parameters $\theta$;}
        \FORALL{$(x,y)$ in the training dataset $S$}
        \STATE{generated $x^{adv}$ with originally fixed $\epsilon$;}
        \STATE{$\Delta{L_{avg}} = L(\theta,x^{adv} ,y) - L(\theta,x,y)$;}
            \FORALL{$(x_{i},y_{i})$ and $(x^{adv}_{i},y_{i})$ in batch}
            \STATE{$\Delta{L} =  L(\theta,{x^{adv}_{i}} ,{y_i}) - L(\theta,{x_i},{y_i})$;}
            \STATE{$\gamma  = \epsilon  \cdot (\left| {\Delta L - \Delta {L_{avg}}} \right|/\Delta {L_{avg}})$;}
            \IF{$\Delta L>\Delta{L_{avg}}$}
            \STATE{${\epsilon _{DPAAT}} = {\left\| {{x^{adv}_{i}} - {x_i}} \right\|_p} - {\gamma }$;}
            \ELSE
            \STATE{${\epsilon _{DPAAT}} = {\left\| {{x^{adv}_{i}} - {x_i}} \right\|_p} + {\gamma }$;}
            \ENDIF
            \STATE{returned new $x^{adv}$ with $\epsilon_{DPAAT}$ ;}
            \ENDFOR
        \STATE{updated the $\theta$ with the new $x^{adv}$ and $L_{DPAAT}$ in \eqref{Eq12}}
        \ENDFOR

	\end{algorithmic}
\end{algorithm}

Fig. \ref{fig:2-fra} illustrated the framework of the DPAAT together with three major conventional AT methods described in previous sections. The DPAAT framework contained two parts: $1)$ $x$ and $x^{adv}$ were separately injected into a CNN classifier to obtain predicted output labels. Then, the classification loss and adversarial loss were calculated and passed forward to dynamically update $\Delta L$ and $\Delta L_{avg}$; $2)$ After forward calculation, new $x^{adv}$ were put into $L_{DPAAT}$ to update the network parameters $\theta$ of the CNN classifier.

Comparing to conventional AT methods, i.e., AT, SAT, and AMAT, the DPAAT had following advantages: $1)$ \textbf{Dynamic perturbation adaption}: the recognition criterion $\Delta L>\Delta L_{avg}$ was dynamically updated through loss information collection during training; $2)$ \textbf{Lower synchronization loss}: the total loss function $L_{DPAAT}$ guided the DL models to obtain lower synchronization loss; $3)$ \textbf{Deeper exploration for $\textbf{x}$}: both the classification loss and the synchronization loss from $L_{SYN}$ took $x$ into consideration during the training.
\section{Experimental results and analysis}
\label{Section 4}
\subsection{Dataset and Training Details.}
We validated robustness and generalization performances of the DPAAT on the MIC, using dermatology HAM10000 dataset \cite{tschandl2018ham10000}, due to its relatively large size and high lesion diversity, i.e., 7 lesions with a total of 10015 
images, including three lesions of Melanoma, Benign Keratosis, and Melanocytic Nevi. The training and testing dataset contained 800 and 299 lesion images for each class, respectively, referring to \cite{margeloiu2020improving}. To avoid exploding gradients, fully connected layers were first trained for 10 iterations alone. Then, all layers were released to train for 50 iterations with a learning rate of 0.0003, mini-batch size of 32, and an Adam optimizer. Moreover, we set an early stopping mechanism of 5 iterations on the validation dataset to prevent overfitting. All models in this paper were trained with 0.5 of $\alpha$ and 7-step PGD training adversary, with $l_{2}$-norm, $\epsilon=0.3$, and step size of 0.15.
\subsection{Improvements of Robustness and Generalization}
\label{section 4.2}
\begin{table*}[!ht]
\setlength{\tabcolsep}{12pt}
\renewcommand{\arraystretch}{0.85}
\caption{The robustness and generalization results of three widely used CNNs, three commonly used lightweight CNNs, and all six CNNs trained by 7-PGD adversary, under iterative IFGSM and PGD attacks. The best results among AT methods were recorded in bold. The training adversary was 7-PGD with $l_{2}$-norm and $\epsilon=0.3$. \textbf{k-} represented \textbf{k-step}.}
\footnotesize
\centering
\begin{tabular}{@{}llcccccccccccc@{}}
\hline
\hline 

\multicolumn{1}{l}{}                            
& \multicolumn{1}{l}{}    
& \multicolumn{10}{c}{\emph{RAcc} (\%)}
& \multicolumn{2}{c}{\emph{GAcc(\%)}}

\\ 
 \multicolumn{1}{c}{CNNs}         
& \multicolumn{1}{c}{Training Methods} 
                                                
& \multicolumn{2}{c}{FGSM}                     
& \multicolumn{2}{c}{10-IFGSM} 
& \multicolumn{2}{c}{20-IFGSM}
& \multicolumn{2}{c}{20-PGD}                     
& \multicolumn{2}{c}{50-PGD}
& \multicolumn{2}{c}{}  
\\ \hline

\multicolumn{1}{c}{}   
& \multicolumn{1}{l}{\emph{STD}} 
&\multicolumn{2}{c}{59.04} 
&\multicolumn{2}{c}{1.06} 
&\multicolumn{2}{c}{0.04} 
&\multicolumn{2}{c}{12.00}
& \multicolumn{2}{c}{11.60} 
& \multicolumn{2}{c}{75.81}

\\

\multicolumn{1}{c}{}                             
& \multicolumn{1}{l}{\emph{AT}}                                         
&\multicolumn{2}{c}{69.12} 
&\multicolumn{2}{c}{59.84}
&\multicolumn{2}{c}{53.95} 
&  \multicolumn{2}{c}{63.14} 
&  \multicolumn{2}{c}{63.14} 
&  \multicolumn{2}{c}{70.41} 

\\

\multicolumn{1}{c}{\emph{Three widely used CNNs}}                              
& \multicolumn{1}{l}{\emph{SAT}}
&\multicolumn{2}{c}{70.19} 
&\multicolumn{2}{c}{56.32}
&\multicolumn{2}{c}{45.87}
&  \multicolumn{2}{c}{60.19} 
&  \multicolumn{2}{c}{60.19} 
&  \multicolumn{2}{c}{71.66}

\\

\multicolumn{1}{c}{}                              
& \multicolumn{1}{l}{\emph{AMAT}}                           
&\multicolumn{2}{c}{69.83}  
&\multicolumn{2}{c}{56.58} 
&\multicolumn{2}{c}{48.55} 
&  \multicolumn{2}{c}{61.58}  
&  \multicolumn{2}{c}{61.58} 
&  \multicolumn{2}{c}{71.45} 
 
\\

\multicolumn{1}{c}{}                            
& \multicolumn{1}{l}{\emph{DPAAT}}            
&\multicolumn{2}{c}{\cellcolor{black!15}\textbf{71.00}}
&\multicolumn{2}{c}{\cellcolor{black!15}\textbf{60.82}}
&\multicolumn{2}{c}{\cellcolor{black!15}\textbf{54.00}}
&  \multicolumn{2}{c}{\cellcolor{black!15}\textbf{65.46}}
&  \multicolumn{2}{c}{\cellcolor{black!15}\textbf{65.46}}
&  \multicolumn{2}{c}{\cellcolor{black!15}\textbf{71.93}} 

\\\hline

\multicolumn{1}{c}{}   
& \multicolumn{1}{l}{\emph{STD}}  
&\multicolumn{2}{c}{51.27}
&\multicolumn{2}{c}{3.66}  
&\multicolumn{2}{c}{0.76}
&  \multicolumn{2}{c}{12.56}
&  \multicolumn{2}{c}{12.71}
&  \multicolumn{2}{c}{70.77}
\\

\multicolumn{1}{c}{\emph{Three commonly used} }                            
& \multicolumn{1}{l}{\emph{AT}}                                          
&\multicolumn{2}{c}{66.93}
&\multicolumn{2}{c}{55.91}
&\multicolumn{2}{c}{49.18} 
&  \multicolumn{2}{c}{61.35}
&  \multicolumn{2}{c}{61.35}
&  \multicolumn{2}{c}{67.83} 
\\

\multicolumn{1}{c}{\emph{lightweight CNNs}}                              
& \multicolumn{1}{l}{\emph{SAT}}
&\multicolumn{2}{c}{66.57} 
&\multicolumn{2}{c}{51.67}
&\multicolumn{2}{c}{44.76}
&  \multicolumn{2}{c}{58.23} 
&  \multicolumn{2}{c}{58.23}
&  \multicolumn{2}{c}{68.72}

\\

\multicolumn{1}{c}{}                              
& \multicolumn{1}{l}{\emph{AMAT}}                           
&\multicolumn{2}{c}{66.73}  
&\multicolumn{2}{c}{54.93} 
&\multicolumn{2}{c}{47.88} 
&  \multicolumn{2}{c}{60.42}  
&  \multicolumn{2}{c}{60.42} 
&  \multicolumn{2}{c}{68.01} 
 
\\
\multicolumn{1}{c}{}                            
& \multicolumn{1}{l}{\emph{DPAAT} }                
&\multicolumn{2}{c}{\cellcolor{black!15}\textbf{68.63}}
&\multicolumn{2}{c}{\cellcolor{black!15}\textbf{59.79}} 
&\multicolumn{2}{c}{\cellcolor{black!15}\textbf{54.84}} 
&  \multicolumn{2}{c}{\cellcolor{black!15}\textbf{63.81}}
&  \multicolumn{2}{c}{\cellcolor{black!15}\textbf{63.81}}
&  \multicolumn{2}{c}{\cellcolor{black!15}\textbf{69.48}}\\
\hline

 \multicolumn{1}{c}{}   
& \multicolumn{1}{l}{\emph{STD}} 
&\multicolumn{2}{c}{55.15}
&\multicolumn{2}{c}{2.36} 
&\multicolumn{2}{c}{0.41}
&  \multicolumn{2}{c}{12.34}
&  \multicolumn{2}{c}{12.20}  
&  \multicolumn{2}{c}{73.96}
\\

 \multicolumn{1}{c}{}                             
& \multicolumn{1}{l}{\emph{AT}}               
&\multicolumn{2}{c}{68.03}
&\multicolumn{2}{c}{56.56}
&\multicolumn{2}{c}{51.56}
&  \multicolumn{2}{c}{{62.25}}
&  \multicolumn{2}{c}{{62.25}} 
&  \multicolumn{2}{c}{69.12} 
\\

\multicolumn{1}{c}{\emph{All six CNNs}}                              
& \multicolumn{1}{l}{\emph{SAT}}
&\multicolumn{2}{c}{68.38} 
&\multicolumn{2}{c}{54.00}
&\multicolumn{2}{c}{44.09}
&  \multicolumn{2}{c}{59.80} 
&  \multicolumn{2}{c}{59.80}
&  \multicolumn{2}{c}{70.19}

\\

 \multicolumn{1}{c}{}                              
& \multicolumn{1}{l}{\emph{AMAT} }                          
&\multicolumn{2}{c}{68.28}  
&\multicolumn{2}{c}{55.75} 
&\multicolumn{2}{c}{48.22} 
&  \multicolumn{2}{c}{61.71}  
&  \multicolumn{2}{c}{61.84} 
&  \multicolumn{2}{c}{{69.73}} 
 
\\

 \multicolumn{1}{c}{}                            
& \multicolumn{1}{l}{\emph{DPAAT}}               
&\multicolumn{2}{c}{\cellcolor{black!15}\textbf{69.81}}
&\multicolumn{2}{c}{\cellcolor{black!15}\textbf{60.31}} 
&\multicolumn{2}{c}{\cellcolor{black!15}\textbf{54.42}} 
&  \multicolumn{2}{c}{\cellcolor{black!15}\textbf{64.70}}
&  \multicolumn{2}{c}{\cellcolor{black!15}\textbf{64.70}}
&  \multicolumn{2}{c}{\cellcolor{black!15}\textbf{70.71}} \\
\hline
\hline
\end{tabular}
\label{table1}
\end{table*}
In this section, we conducted comprehensive experiments on three widely used CNNs, i.e., ResNet34, InceptionV3, and DenseNet121, and three commonly used lightweight CNNs, i.e., MobileNetV2, ShufflenetV2, and SqueezeNet, to verify effectiveness of the DPAAT. The generalization accuracy (GAcc) was defined as follows \cite{ma2022adaptive},
\begin{align}
    GAcc = ({TP + TN})/({TP + TN + FP + FN}),
\end{align}
where $TP$, $FP$, $FN$, and $TN$ represented true positive, false positive, false negative, and true negative predictions on the testing dataset, respectively. Meanwhile, the testing dataset was crafted into an adversarial dataset containing $x^{adv}$, and Robustness accuracy (RAcc) represented the accuracy rate of the DL models on this adversarial dataset.

Table \ref{table1} showed robustness and generalization results of all six CNNs using PGD and IFGSM, due to their wide applications in previous works \cite{madry2017towards},\cite{bai2020adversarial}. For most of the cases, CNNs trained with the DPAAT gave the best robustness and generalization results, compared to those trained with other adversarial methods, i.e., AT, SAT, and AMAT.\\ 
\indent Specifically, the average robustness of all six CNNs using the DPAAT was improved by 1.78\%, 1.43\%, and 1.53\%, 3.75\%, 6.31\%, and 4.56\%, and 2.86\%, 10.33\%, and 6.20\%, under single-step, 10-step, and 20-step IFGSM attacks, respectively, comparing to those of the AT, SAT, and AMAT as shown in Table \ref{table1}. Moreover, the DPAAT also gave the highest robustness compared to other AT methods when malicious PGD attacked selected networks, wherein the average robustness of all six CNNs using the DPAAT was improved by 2.49\% and 2.76\%, 4.93\% and 4.93\%, and 3.03\% and 2.96\% under 20-step and 50-step PGD attacks, respectively, compared to those of the typical AT, SAT, and AMAT. All these results indicated that the DPAAT achieved superior robustness improvements over major conventional AT methods. This was due to the fact that expansion and shrinkage of the adversarial example space were greatly anchored by static constraints in conventional AT methods, wherein loss changes were ignored and possible attempts beyond static constraints were hindered. Instead, the DPAAT gave the dynamic perturbation features \cite{lei2022semi} by collecting and learning loss information in \eqref{Eq15}, and tried more perturbation sizes in its space, especially for relatively stable data. This allowed perturbation injection to approach the maximum perturbations, and the robustness improvements were greatly facilitated.\\
\indent Although the DPAAT showed superior robustness, significant generalization decline could occur in the MIC. Therefore, we still needed to carefully examine the generalization results of all six CNNs. As shown in Table \ref{table1}, three widely used CNNs with the DPAAT achieved 71.93\% generalization on average, the closest to the STD (75.81\% on average), compared to other AT methods. Meanwhile, the DPAAT had similar effects of generalization improvement on three commonly used lightweight CNNs, where the average generalization (69.48\%) only decreased by 1.29\% compared to the STD (70.77\%).\\
\indent More specifically, the generalization of the DPAAT with adaptive perturbation size was improved by 1.58\%, 0.98\%, and 0.52\% on average, compared to the AT, SAT, and AMAT, respectively, as shown in Table \ref{table1}. Although the DPAAT was similar to the AMAT in terms of shrinking the perturbation sizes for the relatively fragile data, it could dynamically search more perturbation sizes and allowed the DL models to obtain more powerful generalization preservation ability when replacing artificial threshold extractions of perturbation adaptation with automatic updates, especially for relatively stable data.\\
\indent In addition, dynamically updated recognition criterion $\Delta L>\Delta L_{avg}$ in the DPAAT provided a dynamic environment for the whole training. This practically satisfied Curriculum Learning \cite{wang2021survey}, indicating that the dynamic environment brought two advantages for training, i.e., efficiency processing of training data was higher in an early stage, and it could guide the training towards a better local optimum and achieve superior generalization in turn.\\
\indent Moreover, with iterative steps increased from 20 to 50, testings with the PGD methods showed no change of robustness for all AT methods, as shown in Table \ref{table1}, since the ability to defend malicious k-PGD attacks was more powerful for the CNNs. However, robustness decreased significantly with increasing iterative steps when using IFGSM attacks. This could attribute to the fact that $l_{\infty}$-norm distance computation in IFGSM might ignore possible inner structures and patterns of adversarial perturbations \cite{10.1007/978-3-030-59719-1_67}. 

\subsection{Improvements of mean average precision (mAP) and mean average robustness precision (mARP)}
\label{Section4.3}
Mean average precision (mAP) could express prediction results more precisely over accuracy rate and was often used as a critical metric to evaluate performance of the DL models. In this paper, we utilized mAP to assess the synchronization of various AT methods, i.e., 
\begin{align}
\label{Eq18}
&mAP=\frac{1}{{{N_1}}}\sum\limits_{{n_1}}^{{N_1}} {(\frac{1}{{{N_2}}}\sum\limits_{{n_2}}^{{N_2}} {({{(TP/(TP + FP))}_{{n_2}}})} )} \nonumber\\&{({(TP/(TP + FN))_{{n_2} + 1}} - {(TP/(TP + FN))_{{n_2}}})_{{n_1}}},  
\end{align}
where $N_1$ and $N_2$ represented numbers of categories and numbers of data in a category, respectively, and mean average robustness precision (mARP) was similar to the RAcc definition and represented mean average precision on adversarial dataset crafted from the testing dataset.
\begin{table*}[!t]
\setlength{\tabcolsep}{14pt}
\small
\renewcommand{\arraystretch}{0.8}
\caption{Mean Average Precision(mAP), Mean Average Robustness Precision(mARP), Precision, Recall, and F1-score results of six CNNs trained by 7-step PGD with $l_{2}$-norm and $\epsilon=0.3$. The results were obtained under 20-step PGD attacks.}
\centering
\begin{tabular}{@{}llccccc@{}}
\hline\hline
\multicolumn{1}{c}{CNNs}                            
& \multicolumn{1}{l}{Training Methods}                                          & \multicolumn{1}{c}{mAP}                           
& \multicolumn{1}{c}{mARP}             
& \multicolumn{1}{c}{Precision}                           
& \multicolumn{1}{c}{Recall}                      
&\multicolumn{1}{c}{F1-score}
\\ \hline

\multicolumn{1}{l}{}                             
& \multicolumn{1}{l}{\emph{AT}}                            
&\multicolumn{1}{c}{0.6000} 
&  \multicolumn{1}{c}{0.6267}  
&  \multicolumn{1}{c}{0.7109} 
&  \multicolumn{1}{c}{0.7096} 
&  \multicolumn{1}{c}{0.7092} 
\\

\multicolumn{1}{c}{\emph{Three widely used CNNs}}                              
& \multicolumn{1}{l}{\emph{SAT}}
&  \multicolumn{1}{c}{0.6600} 
&  \multicolumn{1}{c}{0.5833}  
&  \multicolumn{1}{c}{0.7195} 
&  \multicolumn{1}{c}{0.7197} 
&  \multicolumn{1}{c}{0.7162} 
\\

\multicolumn{1}{l}{}                              
& \multicolumn{1}{l}{\emph{AMAT}}                           
&  \multicolumn{1}{c}{0.5633}  
&  \multicolumn{1}{c}{0.4867} 
 &  \multicolumn{1}{c}{0.7178} 
&  \multicolumn{1}{c}{0.7152} 
&  \multicolumn{1}{c}{0.7130} 
\\

\multicolumn{1}{l}{}                            
& \multicolumn{1}{l}{\emph{DPAAT}}              
&  \multicolumn{1}{c}{\cellcolor{black!15}\textbf{0.8133}}
&  \multicolumn{1}{c}{\cellcolor{black!15}\textbf{0.7633}}  
&  \multicolumn{1}{c}{\cellcolor{black!15}\textbf{0.7304}} 
&  \multicolumn{1}{c}{\cellcolor{black!15}\textbf{0.7310}} 
&  \multicolumn{1}{c}{\cellcolor{black!15}\textbf{0.7270}} 

\\ \hline

\multicolumn{1}{c}{}                            
& \multicolumn{1}{l}{\emph{AT}}                                          
&  \multicolumn{1}{c}{0.7500}
&  \multicolumn{1}{c}{0.7267}
&  \multicolumn{1}{c}{0.6911} 
&  \multicolumn{1}{c}{0.6890} 
&  \multicolumn{1}{c}{0.6891} 
\\

\multicolumn{1}{c}{\emph{Three commonly used}}                              
& \multicolumn{1}{l}{\emph{SAT}}
&  \multicolumn{1}{c}{0.8100} 
&  \multicolumn{1}{c}{0.7600}
&  \multicolumn{1}{c}{0.7011} 
&  \multicolumn{1}{c}{0.7008} 
&  \multicolumn{1}{c}{0.7001} 

\\

\multicolumn{1}{c}{\emph{lightweight CNNs}}                              
& \multicolumn{1}{l}{\emph{AMAT}}                           
&  \multicolumn{1}{c}{0.7167}  
&  \multicolumn{1}{c}{0.6733} 
&  \multicolumn{1}{c}{0.6958} 
&  \multicolumn{1}{c}{0.6902} 
&  \multicolumn{1}{c}{0.6900} 
 
\\

\multicolumn{1}{l}{}                            
& \multicolumn{1}{l}{\emph{DPAAT} }               
&  \multicolumn{1}{c}{\cellcolor{black!15}\textbf{0.8500}}
&  \multicolumn{1}{c}{\cellcolor{black!15}\textbf{0.8333}}
&  \multicolumn{1}{c}{\cellcolor{black!15}\textbf{0.7082}} 
&  \multicolumn{1}{c}{\cellcolor{black!15}\textbf{0.7080}} 
&  \multicolumn{1}{c}{\cellcolor{black!15}\textbf{0.7049}} 
\\\hline

\multicolumn{1}{l}{}                             
& \multicolumn{1}{l}{\emph{AT}}               
& \multicolumn{1}{c}{{0.6750}}
&  \multicolumn{1}{c}{0.6150}
&  \multicolumn{1}{c}{0.7010} 
&  \multicolumn{1}{c}{0.6993} 
&  \multicolumn{1}{c}{0.6992}
\\

\multicolumn{1}{c}{\emph{All six CNNs}}                              
& \multicolumn{1}{l}{\emph{SAT}}
&  \multicolumn{1}{c}{0.7350} 
&  \multicolumn{1}{c}{0.6717}
 &  \multicolumn{1}{c}{0.7103} 
&  \multicolumn{1}{c}{0.7103} 
&  \multicolumn{1}{c}{0.7082} 

\\

\multicolumn{1}{l}{}                              
& \multicolumn{1}{l}{\emph{AMAT} }                          
&  \multicolumn{1}{c}{0.6400}  
&  \multicolumn{1}{c}{0.5800} 
&  \multicolumn{1}{c}{0.7068} 
&  \multicolumn{1}{c}{0.7027} 
&  \multicolumn{1}{c}{0.7015} 
 
\\

\multicolumn{1}{l}{}                            
& \multicolumn{1}{l}{\emph{DPAAT}}               
&  \multicolumn{1}{c}{\cellcolor{black!15}\textbf{0.8317}}
&  \multicolumn{1}{c}{\cellcolor{black!15}\textbf{0.8150}}
&  \multicolumn{1}{c}{\cellcolor{black!15}\textbf{0.7193}} 
&  \multicolumn{1}{c}{\cellcolor{black!15}\textbf{0.7195}} 
&  \multicolumn{1}{c}{\cellcolor{black!15}\textbf{0.7160}} 
\\\hline
\hline
\end{tabular}
\label{table2}
\end{table*}
\begin{figure}[!t]
\centering
\subfloat[]{\includegraphics[width=0.46\textwidth]{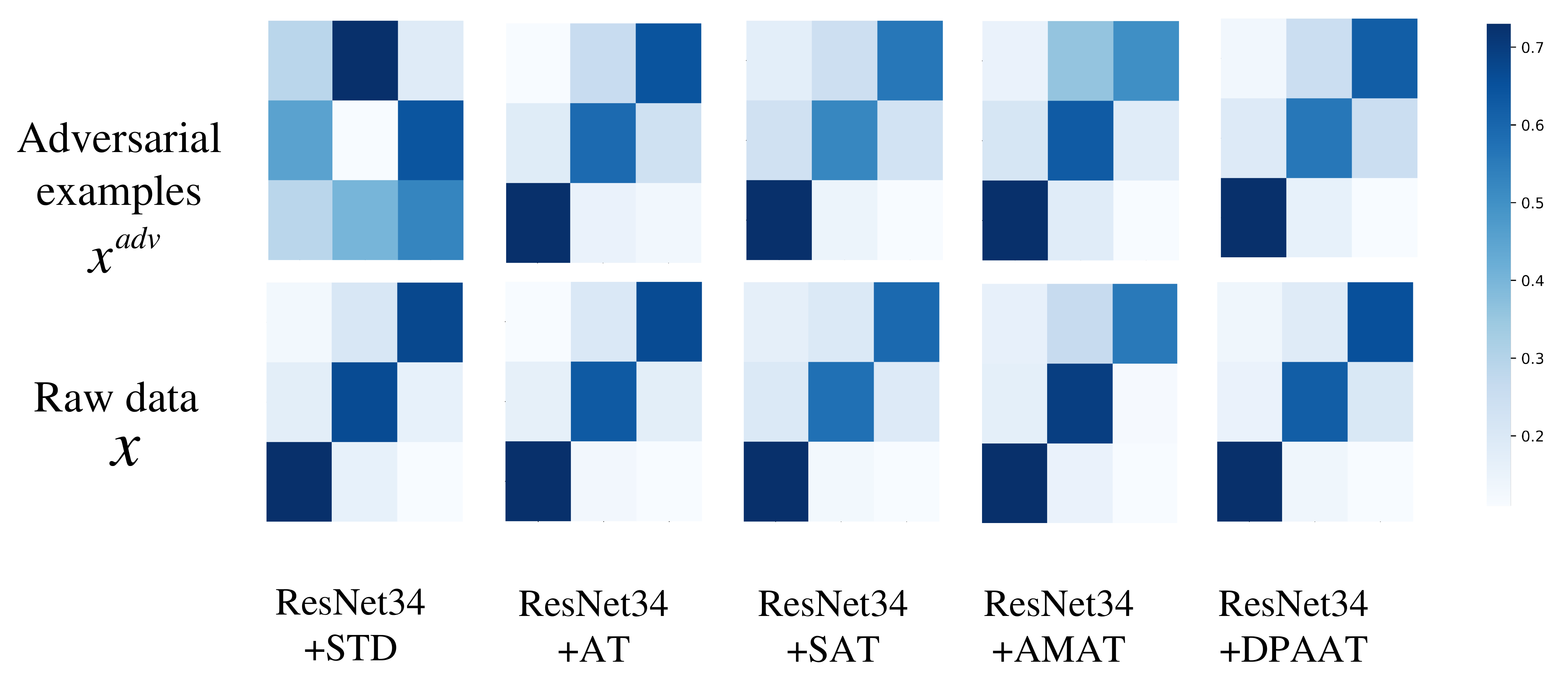}}
\\ \vspace{-0.5cm}
\subfloat[]{\includegraphics[width=0.46\textwidth]{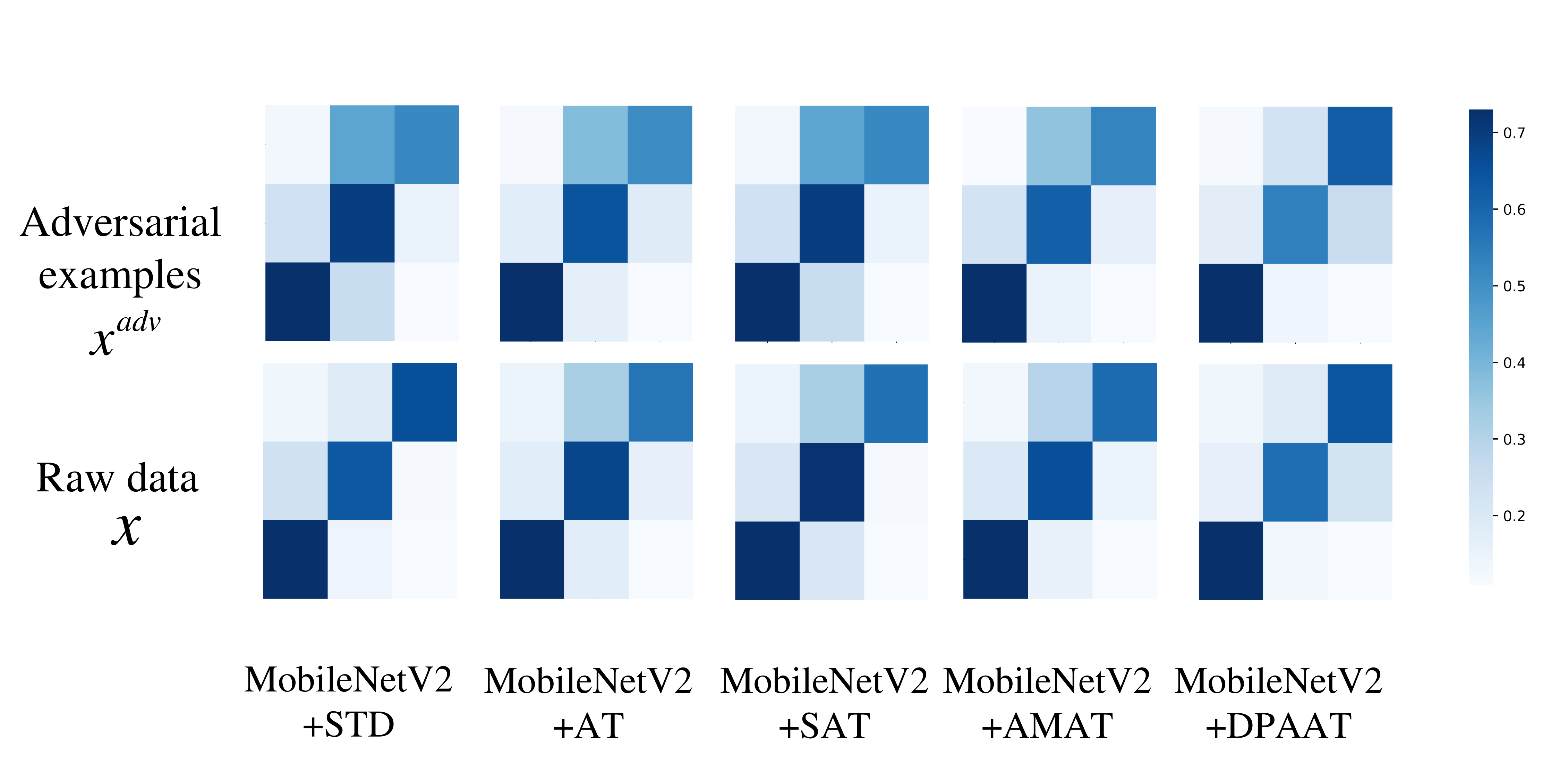}}
\caption{Confusion matrices of ResNet34 and MobileNetV2 on raw data $x$ and adversarial examples $x^{adv}$, wherein $x^{adv}$ were crafted by 20-step PGD attacks. The predicted precision for each class was higher when the colors of the matrices' diagonal were darker.}
\label{fig:3-mat}    
\end{figure}
\\ \indent Obviously, the mAP could be considered to be a more precise expression for generalization for the datasets containing multiple classes, while the corresponding mARP was a precise expression of robustness. It was noticed that both mAP and mARP of the SAT outperformed those of typical AT, indicating that considering the raw data $x$ in the training had a positive impact on performance improvements. Furthermore, although the structure of the loss function of the AMAT originated from that of the SAT, its mAP and mARP were lower, as shown in Table \ref{table2}. In the AMAT, applications of the perturbation adaptation process were restricted without synchronization optimization, further illustrating necessity of the synchronization optimization. As shown in Table \ref{table2}, the mAP and mARP of the DPAAT were improved by 15.67\% and 9.67\%, 19.17\% and 20.00\%, and 14.33\% and 23.50\%, compared to those of the AT, SAT, and AMAT, respectively. This was mainly attributed to the fact that explorations of $x$, in the DPAAT, were much deeper, especially for the explicit application of classification loss $L(\theta, x, y)$ and model logits $y^{ori}$, compared to that of conventional AT methods. In addition, precision, recall, and F1-score of the DPAAT were also improved by 1.83\%, 0.9\%, and 1.25\%, 2.02\%, 0.92\%, and 1.68\%, and 1.68\%, 0.78\%, and 1.45\% on average comparing to those of the AT, SAT, and AMAT, respectively.
\begin{figure}[!ht]
\centering    
\subfloat[]{\includegraphics[width=0.25\textwidth]{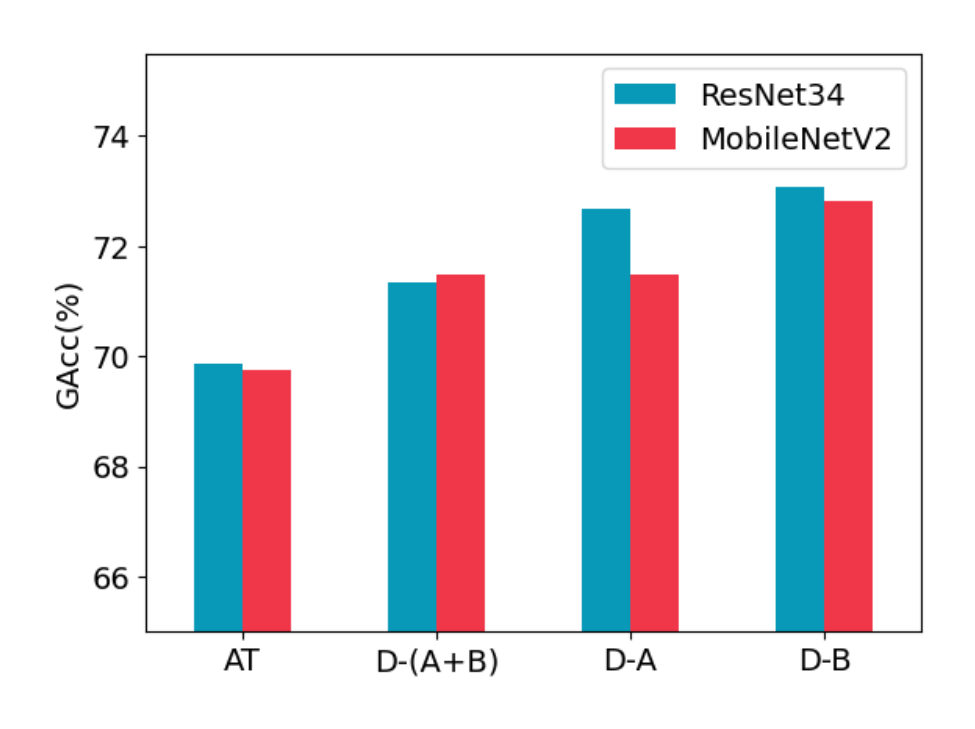}}
\subfloat[]{\includegraphics[width=0.25\textwidth]{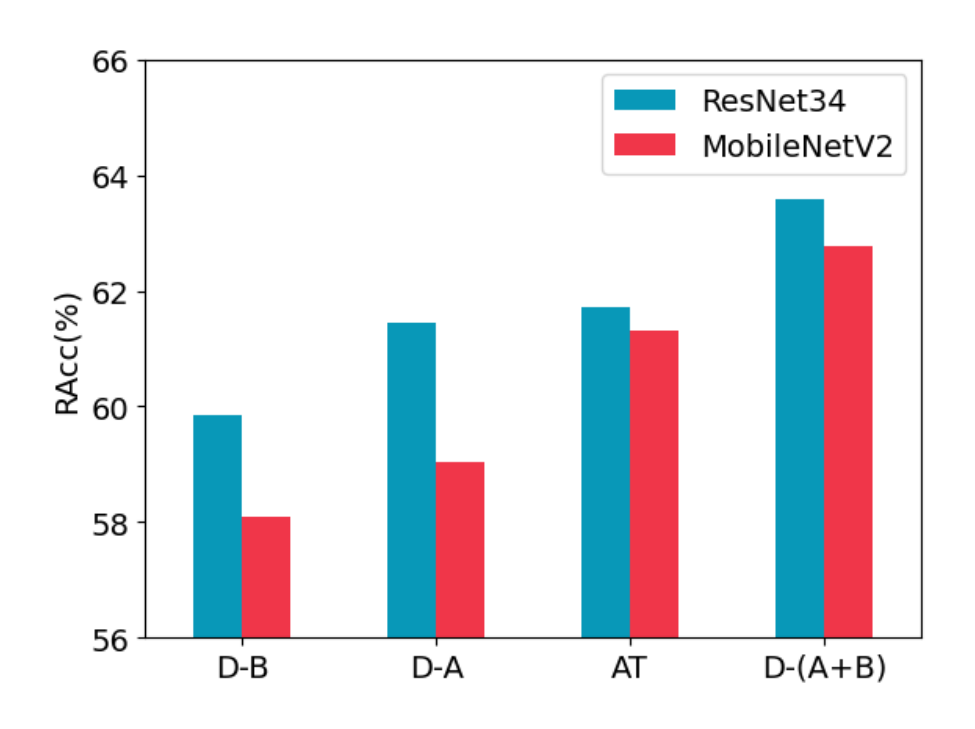}}
\caption{Ablation study for the DPAAT. D-A, D-B, and D-(A+B) represented the DL models implementing only dynamic perturbation adaptation, only synchronization optimization, and both, respectively.}
\label{fig:4-abl}
\end{figure}

In addition, we plotted two types of confusion matrices, with and without malicious attacks for two representative CNNs, i.e., the ResNet34 and MobileNetV2, and did a comparison between the DPAAT and other adversarial methods, including the STD, AT, SAT, and AMAT, as shown in Fig. \ref{fig:3-mat}. Obviously, the STD method behaved badly under malicious attacks, i.e., the diagonal colors were light for $x^{adv}$ of each class. However, the diagonal colors of the confusion matrices using the DPAAT were more even over those of the AT, SAT, and AMAT, for both $x$ and $x^{adv}$. Meanwhile, the grid colors beyond the diagonal in confusion matrices trained with the DPAAT were lighter, indicating that prediction precision of the three lesions became more even under the DPAAT. In fact, improvements of both mAP and mARP and color differences in confusion matrices were reflections of synchronization optimization. 
\subsection{Ablation Studies}
Although the DPAAT could give superior robustness and generalization over conventional AT methods, it was still unclear which of the following two factors contributed more to robustness and generalization improvement, i.e., dynamic perturbation adaptation as described in Section \ref{Section 3.2} or synchronization optimization as in Section \ref{Section 3.1}. Therefore, we did ablation tests on the DPAAT, where the batch size of 64 and training epochs of 100 were selected to train the ResNet34 and MobileNetV2. As shown in Fig. \ref{fig:4-abl}, the generalization of D-A was higher than that of the AT, and the robustness of D-A was similar to that of the AT, illustrating that the dynamic perturbation adaptation of the DPAAT alleviated the generalization decline. Moreover, the generalization of D-B was the highest in the same comparison group, whereas the robustness was lower than that of D-A. The effectiveness of D-B on the robustness became worse when the dynamic perturbation adaptation was not contained in the training process. This elucidated that a fixed perturbation size did not fit the whole adversarial training process and synchronization optimality needed the dynamic perturbation adaptation as a premise to fully give its advantages. When combining the D-A with the D-B, D-(A+B) could achieve the highest robustness and also preserve higher generalization over those using the AT, indicating that the dynamic perturbation adaptation in the DPAAT played a more important role in performance improvements of robustness and generalization.
\subsection{Improvements of Visibility and Interpretability (VI)}
\begin{figure*}[!ht]
    \centering    \includegraphics[width=0.90\textwidth]{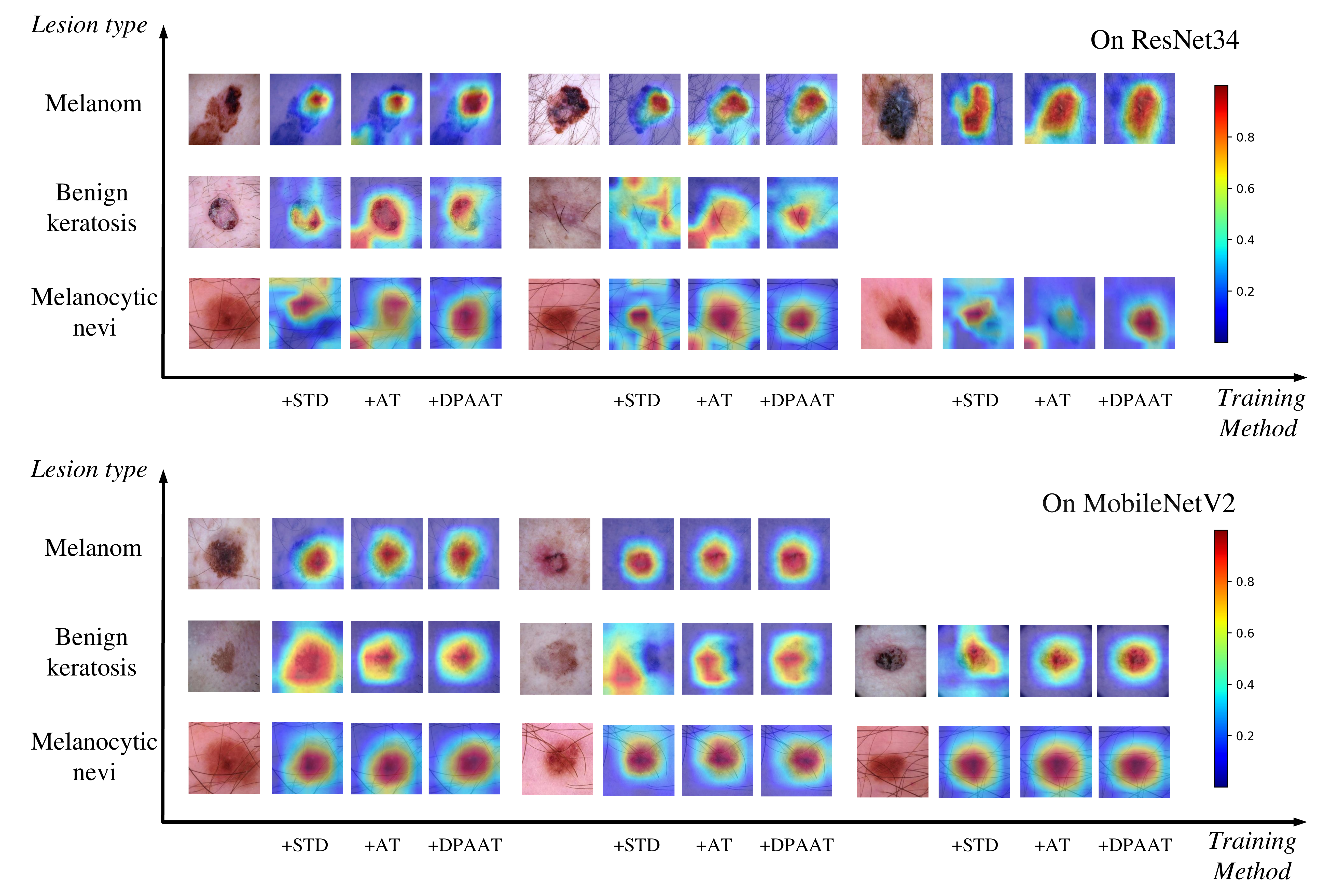}
    \caption{The interpretability visualization of the STD, AT, and DPAAT using the Grad-CAM method. Due to the $ReLU(\cdot)$ activation in \eqref{Eq26}, image pixels showed blue when their contributions to classification result were 0 or negative, while more positive contributions came with a deeper red.}
    \label{fig:5-inter}
\end{figure*}
In previous sections, we proved that the DPAAT established a promising connection between generalization and robustness, making both perform better, comparing to other AT methods. However, improvement of either them, or even both, did not guarantee to bring better VI. In this section, we carefully investigated effects of the DPAAT on VI of the MIC, by carrying out tests on 532 correctly predicted images from the testing medical image dataset. In this paper, the Grad-CAM \cite{selvaraju2017grad}, i.e., a visualization and localization approach for target image, was chosen to illustrate effects of the STD, AT, and DPAAT on VI, due to its applicability to various CNNs and convenience without the need of re-training \cite{selvaraju2017grad}. Specifically,
\begin{align}   
\label{Eq25}
    \alpha _k^c = \frac{1}{Z}\sum\limits_i {\sum\limits_j {\frac{{\delta {y^c}}}{{\delta A_{ij}^k}}} },
\end{align}
where $A$ represented a given convolutional layer, $y^c$ represented target class, $k$ represented feature maps, $i$ and $j$ represented positions in the feature maps, and $\alpha^c_k$ represented an average coefficient contribution of the feature map $k$ to class $c$. The importance score given by the Grad-CAM was the weighted average of the contribution coefficients for the convolution layer, i.e.,
\begin{align}
\label{Eq26}
    Grad - CA{M^c} = {\mathop{\rm Re}\nolimits} LU(\sum\limits_k {\alpha _k^c} {A^k}). 
\end{align}
\indent Fig. \ref{fig:5-inter} showed typical Grad-CAM images of three types of cancers (Melanocytic nevi, Benign keratosis, and Melanoma), using the STD, AT, and DPAAT methods on the ResNet34 and MobileNetV2. Although the Grad-CAM images of the STD and AT had a large amount of redundant and irrelevant regions, i.e., not visually clear compared to corresponding real images, especially for the ResNet34, the DPAAT could cover the lesion regions surprisingly similar to corresponding real images in most of the Grad-CAM images, indicating great improvements of VI for supervised tasks in the MIC. This was critical for understanding the working mechanisms of various CNNs, especially for the MIC tasks.
\section{Conclusion and Future work}
\label{Section 5}
In this paper, we proposed a dynamic perturbation-adaptive adversarial training (DPAAT) method wherein internal dynamic self-transference of loss distribution knowledge was implemented to replace external artificial transference, providing adaptive perturbation sizes to reserve high generalization while improving robustness. Moreover, we used a new loss function to optimize synchronization between robustness and generalization to achieve better performance on both. Comprehensive testing on a typical classification task of the MIC, i.e., screening and classifying on dermatology HAM10000 dataset, showed that the DPAAT not only offered superior robustness and generalization accuracy but also improved interpretability significantly in comparison to several typical AT methods, indicating its great potential as a generic adversarial training method for the MIC.

Although the DPAAT showed superior robustness, generalization, and interpretability performances, future work needed to be done to elucidate effects of adaptive length on the DPAAT. Moreover, testings of the DPAAT on large-scale datasets in other areas using various CNNs were also needed to illustrate its adaptability and usability.

\end{document}